\documentclass[reprint,superscriptaddress,amsmath,amssymb,aps,prb]{revtex4-2}

\usepackage{color}
\usepackage{graphicx}
\usepackage{dcolumn}
\usepackage{bm}
\usepackage{hyperref}
\usepackage{braket}
\usepackage{mathtools}
\usepackage{float}
\usepackage{ulem}

\newcommand{\calE}{\mathcal{E}}

\newcommand{\im}{\mathrm{Im}}

\begin{document}
\preprint{APS/123-QED}

\title{Supercurrent-induced topological phase transitions}

\author{Kazuaki Takasan}
\email{takasan@berkeley.edu}
\affiliation{%
Department of Physics, University of California, Berkeley, California 94720, USA}%
\affiliation{
Materials Sciences Division, Lawrence Berkeley National Laboratory, Berkeley, California 94720, USA
}

\author{Shuntaro Sumita}
\affiliation{%
Condensed Matter Theory Laboratory, RIKEN CPR, Wako, Saitama 351-0198, Japan}%

\author{Youichi Yanase}
\affiliation{%
Department of Physics, Graduate School of Science, Kyoto University, Kyoto 606-8502, Japan}%
\affiliation{
Institute for Molecular Science, Okazaki 444-8585, Japan
}

\date{\today}

\begin{abstract}
We show that finite current in superconductors can induce topological phase transitions, as a result of the deformation of the quasiparticle spectrum by a finite center-of-mass (COM) momentum of the Cooper pairs. To show the wide applicability of this mechanism, we examine the topological properties of three prototypical systems, the Kitaev chain, $s$-wave superconductors, and $d$-wave superconductors. We introduce a finite COM momentum as an external field corresponding to supercurrent and show that all the models exhibit current-induced topological phase transitions. We also discuss the possibility of observing the phase transitions in experiments and the relation to the other finite COM momentum pairing states.
\end{abstract}

\maketitle 

\section{Introduction}
Controlling quantum states of matter is one of the most important subjects in condensed matter physics. For example, pressure or magnetic fields have been widely used. Among them, electric fields, including laser light and accompanying electric currents, are gathering more attention in recent years, thanks to the technological developments for generating electric fields. For instance, driving with AC electric fields (laser light) opens up a new dynamical route to control states of matter, which is called Floquet engineering~\cite{OkaReview2019, RudnerReview2020}. This idea has been applied to a broad range of quantum materials such as topological materials~\cite{Oka2009, Lindner2011, RudnerReview2020}, magnets~\cite{Takayoshi2014, Mentink2015, Sato2016, Claassen2017, Kitamura2017, Takasan2017b}, and superconductors~\cite{Benito2014, Takasan2017a, Babadi2017, Dasari2018, Kennes2019, Chono2020, Dhegani2021}. For DC electric fields,
control of magnetism~\cite{Tokunaga2012, Tokura2014, Takasan2019, Furuya2021} and field-induced superconductivity~\cite{Saito2016_review} has been studied. Electric currents also play an important role in controlling various properties of solids, such as magnetic~\cite{Yanase2014, Zelezny2014, Wadley2016, Bodnar2018, Watanabe2018} and optical~\cite{Aktsipetrov2009, Ruzicka2012, Bykov2012, Khurgin1995, Cheng2014, Takasan2020} properties.

The nature of electric current in superconductors is much different from the one in normal conductors. For example, supercurrent appears in equilibrium and is dissipationless~\cite{Tinkham_book}. Thus, the effect of the supercurrent on the electronic state is expected to be different from the normal metals and may provide another functionality in controlling quantum materials. Indeed, several experiments for the control of superconductors with supercurrent have been reported very recently~\cite{Yang2019, Vaswani2020, Nakamura2020}. In Refs.~\cite{Yang2019, Vaswani2020, Nakamura2020}, after the strong pump by a laser field, the induced supercurrent effectively breaks the inversion symmetry and the second harmonic response is observed in the current-carrying state. This motivates us to study the control of superconductors via electric current~\cite{Doh2006, Moor2017, Volkov2020, Zhu2020, Papaj2021}. From the theoretical point of view, we can study the leading effect of supercurrent within equilibrium states, and it makes the problem more tractable than the normal current, which appears in nonequilibrium states.

Also, the physics related to supercurrent has been gathering renewed attention recently~\cite{Ando2020, Ikeda2020, Yuan2021, Daido2021, He2021, Chazono2021}. For example, the superconducting diode effect, which is the nonreciprocity of critical current, is observed in experiments~\cite{Ando2020} and it stimulates the theoretical studies very recently~\cite{Yuan2021, Daido2021, He2021}. From the viewpoint of device application, superconducting electronics/spintronics, where supercurrent plays a principal role, is becoming an important research field~\cite{Braginski2019, Linder2015, Yang2021, Ikeda2020}. Thus, the importance of studying the effect of current on superconductors is increasing. This is another motivation for this study.

\begin{figure}[t]
\includegraphics[width=6.8cm]{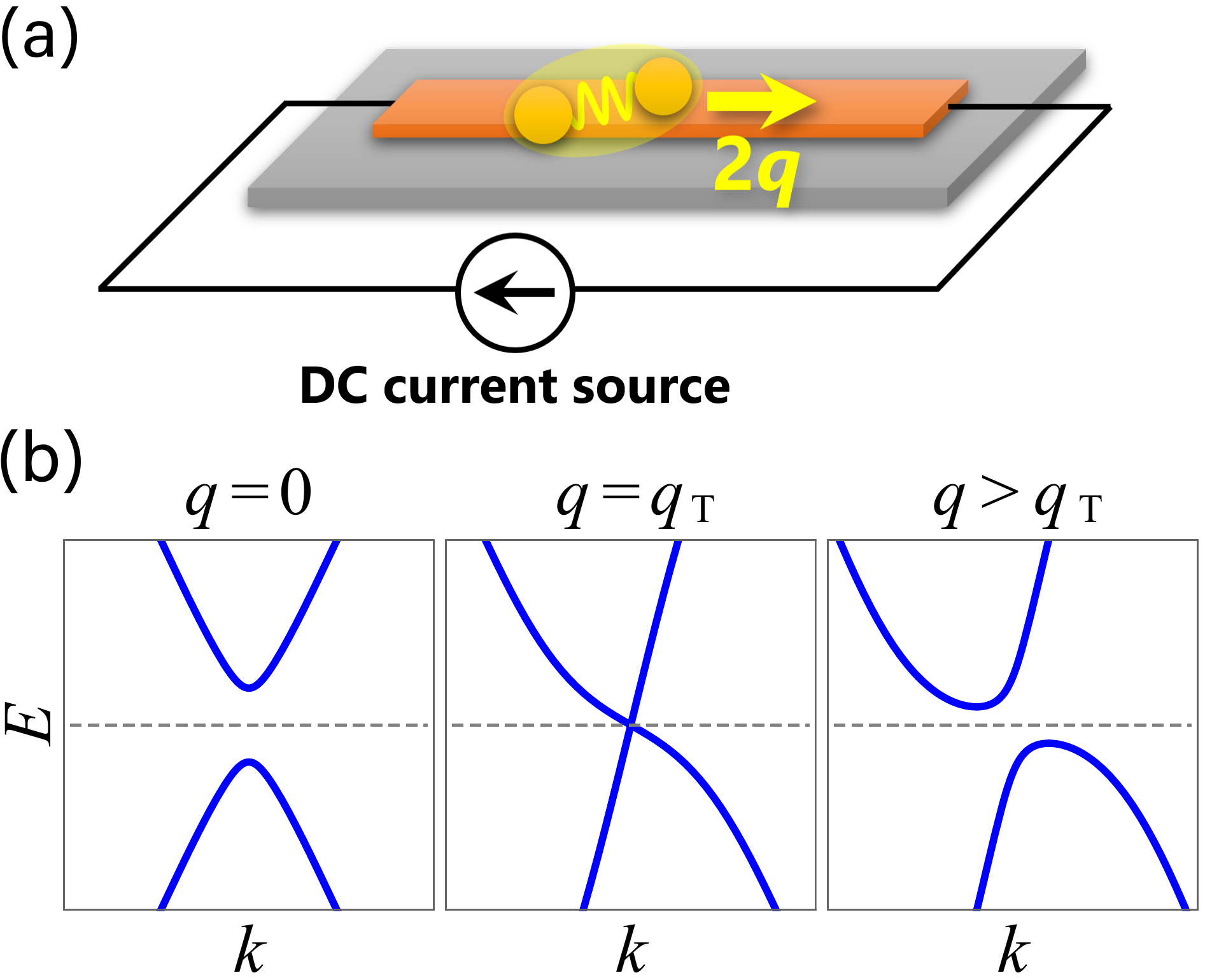}
\caption{
(a) Schematic picture of the setup. Supercurrent is carried by Cooper pairs with the center-of-mass momentum $2 \bm q$.  We consider a narrow strip of superconductors to realize the spatially uniform supercurrent. (b) Quasiparticle spectra in supercurrent-induced topological phase transitions. Supercurrent initially makes the spectrum asymmetric and induces a band touching at $q=q_T$, corresponding to the topological phase transition.
}
\label{fig:fig1}
\end{figure}

In this paper, we study a new possible usage of supercurrent, which is to control the topological phases of matter. 
In a clean superconductor with an external current source shown in Fig.~\ref{fig:fig1}~(a), the Cooper pairs obtain a center-of-mass (COM) momentum $2\bm q$ and carry supercurrent. This results in the modification of the quasiparticle spectrum, as shown in Fig.~\ref{fig:fig1}~(b). It has not been clear whether this modification can realize topological phase transitions. Indeed, as we show later, the supercurrent only gives energy shift in simple single-band models as the leading effect and does not induce any topological phase transitions. However, in this paper, we show that the topological phase transitions occur in a broad range of models with multi-components, such as spin or orbital degrees of freedom. To show that, we study several prototypical models. The first one is the Kitaev chain~\cite{Kitaev2001}, which is a good toy model to demonstrate our idea. The second one is an $s$-wave superconductor with the Rashba spin-orbit coupling (SOC). We show that this model undergoes topological phase transitions induced by supercurrent. The last example is a $d$-wave superconductor with the Rashba SOC. We find that the point nodes are topologically robust even under the supercurrent, although these nodes are known to be gapped out with breaking both inversion and time-reversal symmetries~\cite{Daido2016}. Then, we identify the symmetry protecting the nodes and propose a way to gap out the point nodes with supercurrent. When the symmetry constraint is satisfied, a fully-gapped topological superconductor (TSC) is realized with an infinitesimal supercurrent, which is feasible for the experimental realization.

Before moving to the main part, we mention the previous works related to our results and clarify the difference. First, the superconducting phases studied in this paper are formally the same as the Fulde-Ferrell (FF) states, which is a typical finite COM pairing state~\cite{Fulde1964}. Indeed, topologically non-trivial FF states in ultracold atoms and superconductors have been studied theoretically~\cite{Zhang2013, Qu2013, Hu2019}. The important difference from these studies is that we focus on the systems with an external current source. In other words, the COM momentum of the Cooper pairs is a control parameter in our setup, whereas the thermodynamic stability uniquely determines the COM momentum in the FF states. Therefore our results contain the thermodynamically unstabilizable states without an external current source. Second, Volkov \textit{et al.} studied a supercurrent-induced topological phase transition in a twisted bilayer nodal superconductors very recently~\cite{Volkov2020}. Although we admit their basic idea and the mechanism are similar to ours, there are several important differences as follows: (i) Our focus is on generic phenomena which can occur in various families of superconductors, while they are focusing on the specific interesting example and the relation to the twistronics. (ii) While they consider the out-of-plane current which runs between only the two layers, we consider the in-plane current which runs over the longer length scale. Third, the other related works are proposals of Majorana modes using segmented Fermi surfaces~\cite{Papaj2021, Zhu2020}. In Ref.~\cite{Papaj2021}, the authors considered a finite strip of topological insulators with a proximity-induced superconducting gap. They found that the application of the in-plane magnetic field induces the screening supercurrent, which realizes a topological phase transition. Although this idea is related to our work in the sense that the supercurrent works as the driving force of the phase transition, the topological phase transition in Ref.~\cite{Papaj2021} requires the confinement effect in a finite system.
In contrast, we study whether the topological phase transition occurs or not in a bulk quasiparticle spectrum as an intrinsic effect of supercurrent.

This paper is organized as follows. In Sec.~\ref{sec:theo_descp}, we explain how to describe the current-carrying superconducting state in this study. We also give remarks on the application range of our theory. In Sec.~\ref{sec:Kitaev}, we study the Kitaev chain as one of the simplest examples. In Secs.~\ref{sec:s-wave} and \ref{sec:d-wave}, we investigate two-dimensional $s$-wave and $d$-wave superconductors, respectively. 
We see the quantization of the Berry phase, indicating the topologically non-trivial nature, in Secs.~\ref{sec:s-wave} and \ref{sec:d-wave}, and the origin is explained in Sec.~\ref{sec:quantization}. In Sec.~\ref{sec:discussion}, we discuss the experimental setup to observe the topological phase transitions and the relevance to the other pairing states with finite COM momentum. Finally, we summarize this paper in Sec.~\ref{sec:conclusion}.

\begin{figure*}[t]
\includegraphics[width=15.5cm]{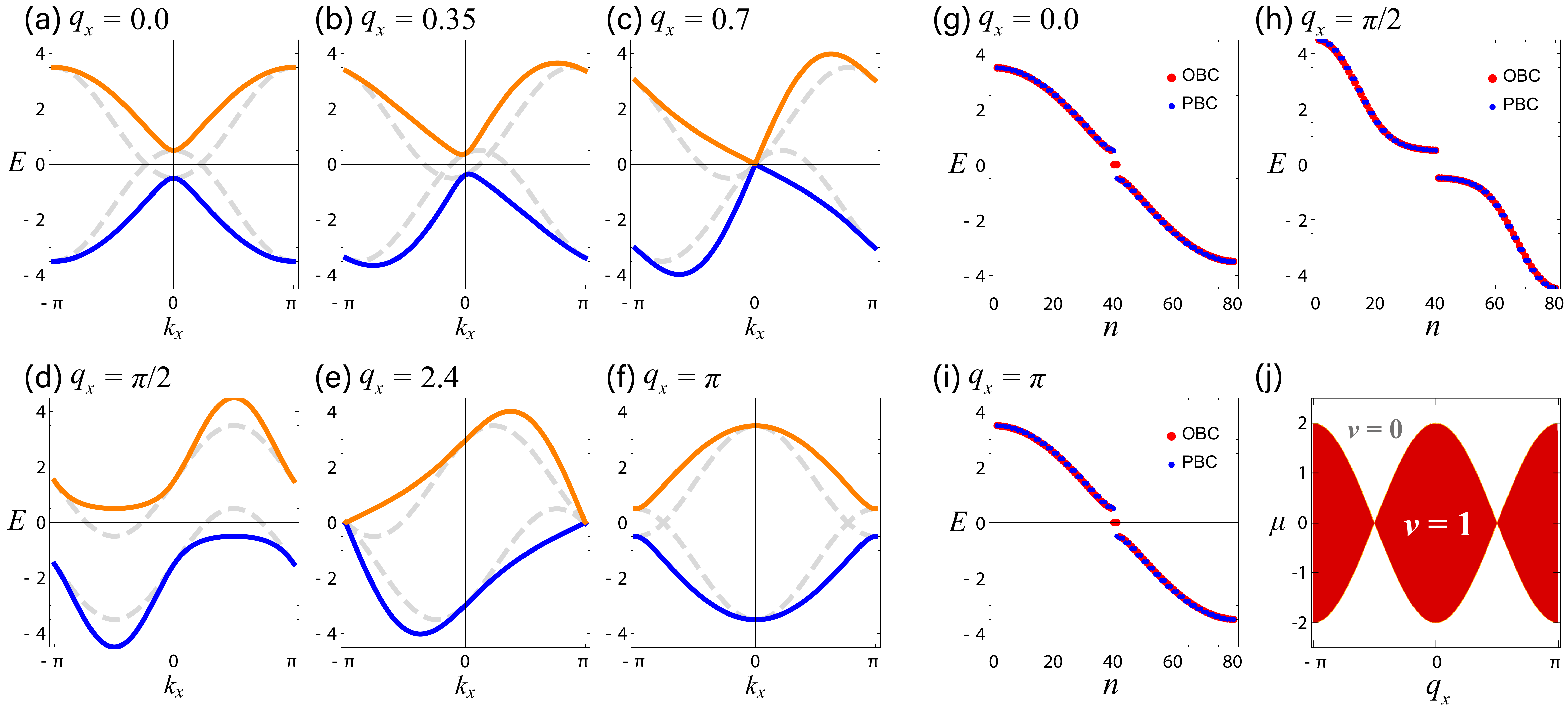}
\caption{Results for the Kitaev chain, Eqs.~\eqref{eq:Kitaev_model_normal} and \eqref{eq:Kitaev_model_OP}. (a)-(f) Quasiparticle spectra for $q_x=0.0, 0.35, 0.7, \pi/2, 2.4, \pi$. (g)-(i) Energy eigenvalues for $q_x=0.0, \pi/2, \pi$ with a finite size (40 sites). The number $n$ denotes the $n$-th eigenvalue from the largest value. The red (blue) symbols correspond to the open (periodic) boundary condition. (f) Topological phase diagram, where the $Z_2$ invariant $\nu$ [Eq.~\eqref{eq:Kitaev_topoinv}] is plotted. Here, hopping amplitude $t$ is used as the energy unit. For (a)-(i), the parameters are set as $\mu=-1.5$ and $\Delta_p=1.0$.
}
\label{fig:Kitaev}
\end{figure*}

\section{Theoretical description of current-driven superconductors}
\label{sec:theo_descp}

In this section, we introduce the basic framework to describe the current-carrying states in superconductors. There are several mechanisms for the generation of electric currents in superconductors. In this study, we consider the electric current carried by the Cooper pairs. Also, we assume that all the Cooper pairs have the same COM momentum $2 \bm q$. This assumption is valid in the narrow strip superconductors, as shown in Fig.~\ref{fig:fig1}~(a), because the current runs near the sample edge due to the Meissner effect~\cite{Tinkham_book}. The detail of the experimental setup will be discussed in Sec.~\ref{sec:discussion_exp}. With the above assumption, the mean-field Hamiltonian with finite supercurrent is given by
\begin{align}
    H&=\sum_{\bm k, s, s^\prime}c^\dagger_{\bm k, s}[H_N(\bm k)]_{s, s^\prime}c_{\bm k, s^\prime}\nonumber\\
    &\quad+\frac{1}{2}\sum_{\bm k, s, s^\prime} \left\{ c^\dagger_{\bm k+ \bm q, s} [\Delta(\bm k)]_{s, s^\prime}  c^\dagger_{-\bm k+ \bm q, s^\prime} + \mathrm{h.c.} \right\},  \label{eq:MFHam_1}
\end{align}
where $c_{\bm k, s}$ ($c^\dagger_{\bm k, s}$) is the annihilation (creation) operator of electrons with momentum $\bm k$ and spin $s$ and $\Delta(\bm k)$ is the superconducting order parameter~\cite{Ikeda2020, Yuan2021, Daido2021, Chazono2021}. While the order parameter is actually determined self-consistently with the gap equation, we do not explicitly solve the gap equation. Instead, we assume specific forms of the gap function for each system. This assumption is expected to be valid with small supercurrent~\footnote{For example, the possibility that another pairing state with finite COM momentum is induced by external current was discussed in Ref.~\cite{Doh2006}. This study shows that the current is needed to be above a certain critical value and thus our assumption is considered to be valid with sufficiently small current.}. The COM momentum $\bm q$ is related to the current density $\bm j$ as $\bm j = e n_s \bm q / m_e$ where $e$, $n_s$, and $m_e$ are the elementary charge, the superfluid density, and the electron mass, respectively. We have to pay attention to that $q$ (=$|\bm q|$) cannot be arbitrary large in experiments because there is an upper bound $q_c$ corresponding to the critical current $j_c$, where the superconducting order breaks down. Thus, our main focus is on the small-$\bm q$ regime, though we will also study the large-$\bm q$ regime to clarify the mathematical structure of the models in the several cases discussed later.

Topological phases in superconductors are characterized by the topologically non-trivial ground state of the Bogoliubov--de~Gennes (BdG) Hamiltonian. The mean-field Hamiltonian~\eqref{eq:MFHam_1} is rewritten into the BdG Hamiltonian,
\begin{align}
    &H=\frac{1}{2} \sum_{\bm k} \bm \Psi^\dagger_{\bm k; \bm q} H_\mathrm{BdG}(\bm k; \bm q) \bm \Psi_{\bm k; \bm q}, \label{eq:MFHam_2} \\
    &H_\mathrm{BdG}(\bm k; \bm q)=
    \begin{pmatrix}
    H_N(\bm k+\bm q) & \Delta(\bm k) \\
    \Delta^\dagger(\bm k) & -H^T_N(-\bm k+\bm q)
    \end{pmatrix},
\end{align}
where $\bm \Psi_{\bm k;\bm q}=(c_{\bm k+\bm q, \uparrow}, c_{\bm k+\bm q, \downarrow}, c^\dagger_{-\bm k+\bm q, \uparrow}, c^\dagger_{-\bm k+\bm q, \downarrow})^T$. 
Diagonalizing this BdG Hamiltonian, we obtain the quasiparticle spectrum under a supercurrent. Below, we study the topological properties of this Hamiltonian for specific examples. 

Before moving to the specific cases, using Eq.~\eqref{eq:MFHam_2}, we reveal that current-induced topological phase transition is difficult to occur in simple single-band models. Let us consider a single-band spin-1/2 system 
with the normal Hamiltonian $H_N(\bm k) = \xi(\bm k) \sigma_0 = \{\varepsilon(\bm k)-\mu\} \sigma_0$, where $\varepsilon(\bm k)$ is a symmetric dispersion, $\mu$ is the chemical potential, and $\sigma_0$ is the identity matrix in the spin space. We do not assume a specific form for the superconducting order parameter $\Delta(\bm k)$. Since $H_N(\pm \bm k+\bm q) = \xi(\bm k)\sigma_0 \pm \bm q \cdot (\partial \xi / \partial \bm k) \sigma_0+O(q^2)$, the BdG Hamiltonian with supercurrent becomes $H_\mathrm{BdG}(\bm k; \bm q)= H_\mathrm{BdG}(\bm k; \bm 0) + \{\bm q \cdot (\partial \xi / \partial \bm k)\} I+O(q^2)$ where $I (=\sigma_0 \otimes \tau_0)$ is the $4 \times 4$ identity matrix with the identity matrix in the Nambu space $\tau_0$. Therefore, the supercurrent only gives the energy shift as the leading effect in this case, and thus topological phase transition does not occur.
This energy shift is known as the Doppler shift~\cite{Matsuda2006}.

\section{Kitaev chain}
\label{sec:Kitaev}

In this section, we consider the current-carrying state in the Kitaev chain~\cite{Kitaev2001}, which is defined with the normal part Hamiltonian and the order parameter,
\begin{align}
H_N(k_x)&=-2t\cos k_x -\mu, \label{eq:Kitaev_model_normal} \\
\Delta(k_x)&=-2\Delta_p i \sin k_x, \label{eq:Kitaev_model_OP}
\end{align}
where $t$, $\mu$, and $\Delta_p$ are the hopping amplitude, the chemical potential and the $p$-wave pairing strength, respectively. Without current, the ground state of this model becomes a TSC for $|\mu/t|<2$ which hosts Majorana end modes~\cite{Kitaev2001, Alicea2012}. 

\begin{figure*}[t]
\includegraphics[width=16cm]{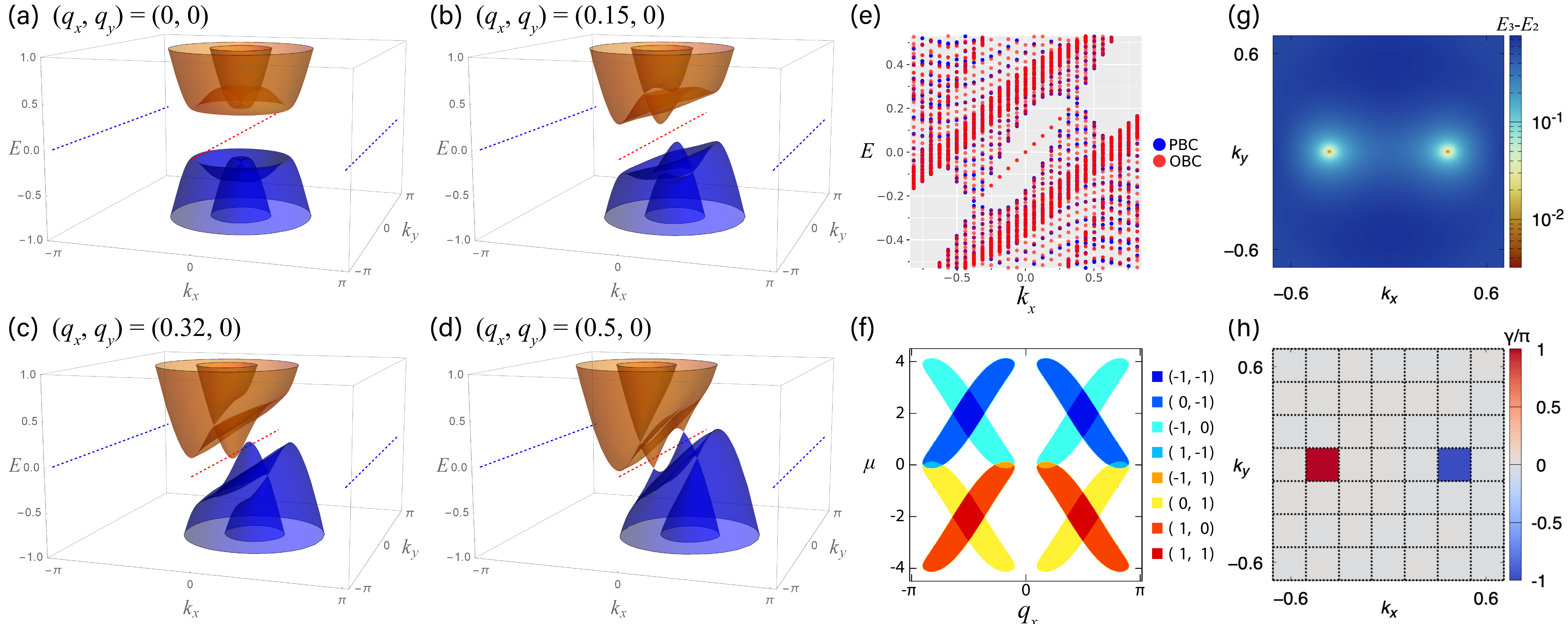}
\caption{
Results for the $s$-wave superconductor, Eqs.~\eqref{eq:swave_model_normal} and \eqref{eq:swave_model_OP}. (a)-(d) Quasiparticle spectra for $\bm q$=(0,0), (0.15,0), (0.32,0), (0.5,0). (e) Quasiparticle spectra for $\bm q=(0.5,0)$ with a finite size only in the $y$-direction (100 sites). The red (blue) symbols correspond to the open (periodic) boundary condition. (f) Topological phase diagram, where the winding number [Eq.~\eqref{eq:wind_num}] is plotted. Different colors correspond to the different sets of ($w(0)$,$w(\pi)$) as shown in the legend. (g) The energy gap $E_3(\bm k)-E_2(\bm k)$ for $\bm q=(0.5,0)$, where $E_n(\bm k)$ is the energy eigenvalues of the BdG Hamiltonian ($n=1,2,3,4$ and $E_n > E_n^\prime$ for $n>n^\prime$). The intense two spots represent the nodes shown in the panel (d). (h) Berry phase [Eq.~\eqref{eq:Berry}] for $\bm q=(0.5,0)$. The plotted value is divided by $\pi$. The integration path is each square denoted by the dotted lines. For all figures, we set the hopping amplitude $t$ as the energy unit and the other parameters are $\mu=-3.8$, $\alpha=1.0$, and $\Delta_s=0.3$.
}
\label{fig:swave}
\end{figure*}

Let us consider the effect of supercurrent. The quasiparticle spectrum with finite $q_x$ is shown in Figs.~\ref{fig:Kitaev}~(a)-(f). With small $q_x$, the spectrum becomes asymmetric respecting the inversion symmetry breaking by the supercurrent [Fig.~\ref{fig:Kitaev}~(b)]. A band touching appears at $q_x \sim 0.7$ and then the gap reopens~[Figs.~\ref{fig:Kitaev}~(c)~and~(d)]. With larger $q_x$, the gap closes at $q_x \sim 2.4$ and then it opens again~[Figs.~\ref{fig:Kitaev}~(e)~and~(f)]. This behavior suggests that topological phase transitions occur two times at the band touching points in this sequence. Indeed, the number of the Majorana end modes is changed at these points as shown in Figs.~\ref{fig:Kitaev}~(g)-(i). This is direct evidence of the topological phase transitions induced by supercurrent.

Next, we clarify the topological index for this model. The BdG Hamiltonian is written as $H_\mathrm{BdG}(\bm k)=h_0(k_x) \tau_0 + \bm h(k_x) \cdot \bm \tau$
with $h_0 = 2t \sin q_x \sin k_x$ and $\bm h = (0, 2 \Delta_p \sin k_x, -2t \cos q_x \cos k_x -\mu)$ where $\bm \tau=(\tau_x, \tau_y, \tau_z)$ denotes the Pauli matrix in the Nambu space. The term proportional to $\sigma_0$ does not affect to the topological properties, and thus the BdG Hamiltonian is essentially the same as the original Kitaev chain with a replacement of $t \to t \cos q_x$. The topological index is given by $\nu=\mathrm{sgn}[h_z(0)]\mathrm{sgn}[h_z(\pi)]$~\cite{Alicea2012} and thus
\begin{align}
    \nu
    &=\mathrm{sgn}(\mu^2-4t^2\cos^2 q_x). \label{eq:Kitaev_topoinv}
\end{align}
Plotting this for the different values of $q_x$ and $\mu$, we obtain the topological phase diagram shown in Fig.~\ref{fig:Kitaev}~(j). We can see that the supercurrent-induced topological phase transition occurs in a broad range of the parameter space. Note that we can use the original $Z_2$ index for the Kitaev chain because the particle-hole (PH) symmetry is only needed to define this index~\cite{Alicea2012} and the PH symmetry is preserved even under supercurrent. 

The phase diagram suggests that the current-induced topological phase transitions only occur in the parameter range where the topologically non-trivial ground state is already realized. This means that it is impossible to induce topologically non-trivial phase starting from the trivial phase. However, this feature depends on the details of models, and we show that it is possible for other models in Secs.~\ref{sec:s-wave}~and~\ref{sec:d-wave}. 
The advantage of the Kitaev chain is that we can easily see the mechanism of supercurrent-induced phase transitions. The bare electron and hole bands are shown with the gray dashed lines in Figs.~\ref{fig:Kitaev}~(a)-(f). The convex downward (upward) curve is the electron (hole) band. The figures clearly represent that the origin of the topological phase transitions is the band inversion driven by the COM momentum $q_x$, which horizontally shifts the electron and hole bands in the opposite directions. While the other examples below are more complex and difficult to see the band inversion directly, the mechanism is the same as in this case.

\section{s-wave superconductors}
\label{sec:s-wave}
In this section, we consider a model for $s$-wave superconductors in two dimensions as a simple but more realistic example. As shown in Fig.~\ref{fig:fig1}~(a), we consider the sample fabricated on a substrate. The heterostructure breaks the inversion symmetry, and the effect is taken into account as the Rashba-type SOC. The model is defined as
\begin{align}
H_N(\bm k)&= \xi(\bm k)\sigma_0+ \alpha \bm g_R (\bm k)\cdot \bm \sigma, \label{eq:swave_model_normal} \\
\Delta(\bm k)&=\Delta_s i\sigma_y, \label{eq:swave_model_OP}
\end{align}
with $\xi(\bm k)=-2t (\cos k_x + \cos k_y) -\mu$ and $\bm g_R(\bm k)=(-\sin k_y, \sin k_x, 0)$. Here, $\bm \sigma=(\sigma_x, \sigma_y, \sigma_z)$ is the Pauli matrices in the spin space.
This model with the Zeeman magnetic field is known as a prototypical model of TSCs~\cite{Sato2009}. 
Since supercurrent breaks the time-reversal symmetry as the Zeeman field does, it is natural to ask if the topological superconductivity is realized with supercurrent or not. As shown below, topologically non-trivial phases are induced by supercurrent, though the phases are different from those in the Zeeman field case.

The quasiparticle spectra under finite current are shown in Figs.~\ref{fig:swave}~(a)-(d). For simplicity, we consider the current parallel to the $x$-axis. First, for small $q \equiv |\bm q|$, the spectrum becomes distorted and  asymmetric~[Fig.~\ref{fig:swave}~(b)]. Then, at $q \sim 0.32$, a band touching occurs~[Fig.~\ref{fig:swave}~(c)]. These behaviors are the same as what we have seen in the Kitaev chain (see the previous section). In contrast, the next step is different from the case of the Kitaev chain, where the system becomes fully gapped again. For $q>0.32$, the quasiparticle spectrum is still gapless, and there exist two gapless points~[Fig.~\ref{fig:swave}~(d)]. These point nodes are robust under small changes of the parameters, and thus these are expected to be topologically protected. Indeed, there appear chiral Majorana edge modes connecting these two point nodes, as shown in the energy spectrum calculated with the open boundary condition in the $y$-direction~[Fig.~\ref{fig:swave}~(e)]. This suggests that the nodes are topologically protected and the system is in a topologically non-trivial phase. The chiral Majorana modes in the different edges propagate in the same direction, as seen from the spectrum shown in Fig.~\ref{fig:swave}~(e). This is known as the unidirectional Majorana state appearing under a tilted magnetic field~\cite{Daido2017}. 
Seeing Fig.~\ref{fig:swave}~(e), the Majorana states are buried in the bulk spectrum due to the finite density of states at the Fermi level, and thus it might seem that the TSC becomes ill-defined and the Majorana modes are not able to be observed. However, the TSC can be defined with topological indices as shown below, and the Majorana state is observable with choosing the adequate sides of the surfaces. We will discuss experimental methods further in Sec.~\ref{sec:discussion_exp}.

Next, we explain the topological index characterizing these gapless points. There exist two topological indices, which are related to each other and essentially have the same information in the limited cases as we mention below. The first one is the winding number. When the current is parallel to the $x$-axis, the following chiral symmetry,
\begin{align}
\Gamma=M_x T C_{ph}, \label{eq:chiralsym}
\end{align}
exists only at $k_x=0$ and $k_x=\pi$. Here, $M_x$, $T$, and $C_{ph}$ are the operators for mirror symmetry in the $x$-direction, time-reversal symmetry, and PH symmetry, respectively. Note that this is different from the widely-used chiral symmetry $T C_{ph}$. In our notation, they are explicitly given as
\begin{gather}
M_x\!=\!\!
\begin{pmatrix}
i \sigma_x & \!\!\!\!\! 0 \\
0 & \!\!\!\!\! - i \sigma_x
\end{pmatrix}, \quad \!\!\!
T\!=\!
\begin{pmatrix}
i \sigma_y &\!\! 0 \\
0 &\!\! i \sigma_y
\end{pmatrix}\!K, \quad \!\!\!
C_{ph}\!=\!
\begin{pmatrix}
0 & \sigma_0 \\
\sigma_0 & 0
\end{pmatrix}\!K,\nonumber
\end{gather}
where $K$ is the complex conjugation.
Using this chiral symmetry $\Gamma$, we can define the winding number as below,
\begin{align}
    w(k_x)=\frac{1}{2\pi}\mathrm{Im}\left[ \int_{-\pi}^{\pi}\!\!dk_y  \partial_{k_y} \! \ln \det \hat{q}(\bm k) \right], \label{eq:wind_num}
\end{align}
where $\hat{q}(\bm k)$ is a $2 \times 2$ matrix defined via
\begin{align}
    U^\dagger_{\Gamma} H_{\mathrm{BdG}}(\bm k) U_{\Gamma} = \begin{pmatrix}
    0 & \hat{q}(\bm k) \\
    \hat{q}^\dagger(\bm k) & 0
    \end{pmatrix},
    \label{eq:off-diagonal_BdG}
\end{align}
and $U_\Gamma$ is a unitary matrix satisfying $U^\dagger_{\Gamma} \Gamma U_{\Gamma}=\mathrm{diag}(1,1,-1,-1)$~\cite{Sato2011}. The integration paths for $k_x=0, \pi$ are shown by red and blue dashed lines respectively in Figs.~\ref{fig:swave}~(a)-(d). We calculate $w(0)$ and $w(\pi)$ for different parameters and the results are summarized as a topological phase diagram in Fig.~\ref{fig:swave}~(f). There appear several distinct phases characterized by the winding number. When $w(0)$ [$w(\pi)$] is nonzero, there appear edge modes around $k_x=0$ [$k_x=\pi$] when we take the open boundary condition in the $y$-direction.

\begin{figure*}[t]
\includegraphics[width=18cm]{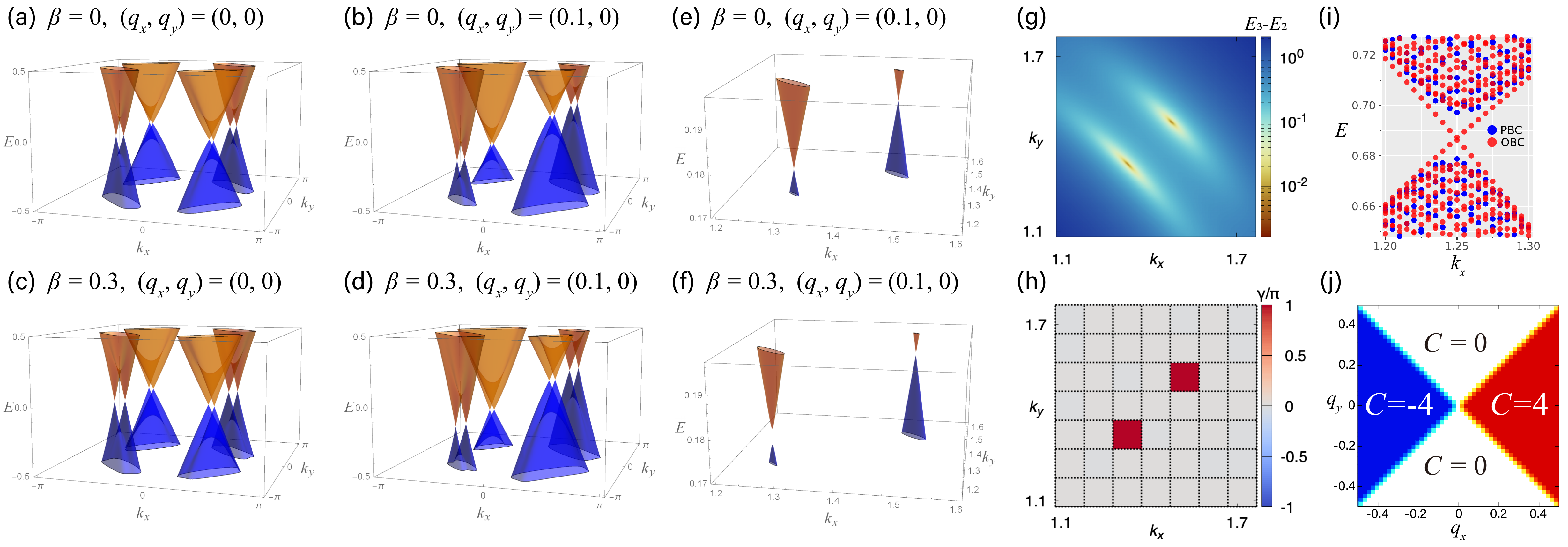}
\caption{
Results for the $d$-wave superconductors, Eqs.~\eqref{eq:dwave_normal} and \eqref{eq:dwave_OP}. 
(a)-(d) Quasiparticle spectra for $\bm q$ = (0, 0), (0.1, 0) and $\beta=$ 0, 0.3. (e) and (f) are the enlarged views of (b) and (d) around the nodes in $k_x, k_y > 0$. (g) The energy gap $E_3(\bm k)-E_2(\bm k)$ for $\bm q=(0.1,0)$ and $\beta=0$, where $E_n(\bm k)$ is the energy eigenvalues of the BdG Hamiltonian ($n=1,2,3,4$ and $E_n > E_n^\prime$ for $n>n^\prime$). The intense two spots represent the nodes shown in the panel (e). (h) Berry phase [Eq.~\eqref{eq:Berry}] for $\bm q=(0.1,0)$ and $\beta=0$. The plotted value is divided by $\pi$. The integration path is each square denoted by the dotted lines. (i) Quasiparticle spectra for $\bm q=(0.4,0)$ and $\beta=0.3$ with a finite size only in the $y$-direction (500 sites). The red (blue) symbols correspond to the open (periodic) boundary condition. (j) Topological phase diagram for $\beta=0.3$, where the Chern number [Eq.~\eqref{eq:Chern}] is plotted. For all figures, we set the hopping amplitude $t$ as the energy unit and the other parameters are $t^\prime=0.2$, $\mu=-0.7$, $\alpha=0.3$, $\Delta_d=0.5$ and $\Delta_p=0.2$. 
}
\label{fig:dwave}
\end{figure*}

The other index is the Berry phase $\gamma$ defined as
\begin{align}
    \gamma = \frac{1}{i} \sum_{n=1,2} \oint_C d k_i  \bra{u_n(\bm k)}\partial_{k_i}\ket{u_n(\bm k)} \quad (\mathrm{mod}~2\pi), \label{eq:Berry}
\end{align}
where $\ket{u_n(\bm k)}$ is the Bloch state and $C$ is a closed path in the Brillouin zone. The bands of $n=1,2$ are taken as the two lowest energy eigenstates for each momentum. We numerically calculated the Berry phase using a method for the discretized Brillouin zone~\cite{Hatsugai2006} and the results are summarized in Fig.~\ref{fig:swave}~(e). This figure shows that the Berry phase takes a non-zero quantized value when the path $C$ encloses the point node. 
This result directly shows that the point nodes are topologically protected. The Berry phase is same as the parity of the winding number when both are well-defined as discussed in Appendix~\ref{sec:wind_Berry}. However, the sharp difference appears when the COM momentum $\bm q$ is not parallel to the $x$-axis or the $y$-axis. In such cases, the chiral symmetry [Eq.~\eqref{eq:chiralsym}] does not exist, and thus we cannot define the winding number. On the other hand, the quantized Berry phase is always well-defined. Indeed, we can confirm the quantization of the Berry phase and the appearance of the edge modes for any current direction. This quantization of the Berry phases is understood from the classification of topological phases. We will discuss this point in Sec.~\ref{sec:quantization}.

\section{d-wave superconductors}
\label{sec:d-wave}

Next, let us consider $d$-wave superconductors as another example. 
The results in this section are important from the viewpoint of experimental realization. In the examples presented so far, the transition momenta $q_T$, where the topological phase transition occurs, are non-zero. As mentioned in Sec.~\ref{sec:theo_descp}, $q$ must be smaller than the critical momentum $q_c$. Thus, $q_T$ also needs to be smaller than $q_c$. This means that we need to find the parameters to make $q_T$ sufficiently small. We can avoid the difficulty if we can find the case where topological phase transitions occur with infinitesimal momentum (i.e., $q_T=0$). Below, we will see that the $d$-wave superconductor with a certain perturbation is such a fascinating case. 

We consider a two-dimensional $d$-wave superconductor on a substrate, whose effect is taken into account as the Rashba SOC as in Sec.~\ref{sec:s-wave}. This can be regarded as a model for cuprate superconductor thin films~\cite{Bollinger2011, Leng2011}. The model is defined as 
\begin{align}
H_N(\bm k)&= \xi(\bm k)\sigma_0+ \alpha \bm g_R (\bm k) \cdot \bm \sigma, \label{eq:dwave_normal}\\ 
\Delta(\bm k)&=\{\psi_d(\bm k)\sigma_0+\bm d(\bm k)\cdot \bm \sigma\}i\sigma_2, \label{eq:dwave_OP}
\end{align}
with
$\xi(\bm k)=-2t (\cos k_x + \cos k_y) + 4 t^\prime \cos k_x  \cos k_y -\mu$ and $\bm g_R(\bm k)=(-\sin k_y, \sin k_x, 0)$. The order parameter consists of the spin-singlet sector $\psi_d(\bm k)= \Delta_d(\cos k_x -\cos k_y)$ and the spin-triplet sector $\bm d(\bm k)= \Delta_p (\sin k_y, \sin k_x, 0)$ where $\Delta_d \gg \Delta_p$. The spin-triplet component is admixed to the dominant $d$-wave component because of the inversion symmetry breaking by the substrate or intrinsic crystal structure~\cite{Daido2016, footnote1}.

As discussed in Ref.~\cite{Daido2016}, breaking both inversion and time-reversal symmetry in this model induces TSCs. Following this idea, it is expected that supercurrent can also induce topological phase transitions from the symmetry viewpoint. However, against this naive guess, we can show that the application of finite current to this model does not cause topological phase transitions. In other words, the gap nodes in $d$-wave superconductors are robust against supercurrent. Indeed, as shown in Figs.~\ref{fig:dwave}~(b) and (e), the quasiparticle spectrum becomes asymmetric but the nodes still exist with a finite $q$. 

The mechanism of the protection can be understood in two ways. One is based on the perturbation theory. As presented in Appendix~\ref{sec:perterbation}, the size of the current-induced energy gap at the nodal points $\bm k_0$, which we denote by $\Delta_\mathrm{node}(\bm q)$, is estimated with the perturbation theory as
\begin{align}
    \Delta_\mathrm{node}(\bm q) = 2 \left| \frac{\bm g^\prime_{\bm q}(\bm k_0) \cdot \hat{\bm g}(\bm k_0)\times \bm d(\bm k_0)}
    {\bm g(\bm k_0)} \right| + O(q^2), \label{eq:current_induced_gap}
\end{align}
where $\bm g(\bm k)=\alpha \bm g_R(\bm k)$, $\hat{\bm g}(\bm k)=\bm g(\bm k)/|\bm g(\bm k)|$ and $\bm g_{\bm q}^\prime(\bm k)=\sum_{i=x,y} q_i (\partial \bm g(\bm k)/\partial k_i)=\alpha (- q_y \cos k_y, q_x \cos k_x, 0)$. In our case, all the vectors $\bm g_{\bm q}^\prime(\bm k)$, $\hat{\bm g}(\bm k)$, and $\bm d(\bm k)$ in Eq.~\eqref{eq:current_induced_gap} are in the $x$-$y$ plane, and thus this value must be zero. Therefore, no energy gap is induced, and the nodes are robust to current. The other approach is based on the band topology. As in the $s$-wave case, we calculate the Berry phase with respect to the path enclosing the nodes and find that this takes a quantized value not changed even under a finite current, as shown in Fig.~\ref{fig:dwave}~(h). This means that the nodes are topologically protected even under finite current.

From the above discussion about the protection of the gap nodes, we notice how to induce a finite gap by supercurrent. From Eq.~\eqref{eq:current_induced_gap}, we see that
the energy gap can be finite when the $g$-vector $\bm g(\bm k)$ has the $z$-component. Also, as we will discuss later in Sec.~\ref{sec:quantization}, for topological protection, the quantization of the Berry phase is protected by symmetry. In our case, a composite symmetry $TC_{2z}$ protects the nodes even under finite current. Here, $T$ and $C_{2z}$ represent the time-reversal symmetry and the two-fold rotational symmetry along the $z$-axis, respectively. Thus, the perturbation breaking this symmetry is expected to violate the topological protection and open a finite gap at the nodal points. Following the above arguments, we consider 
\begin{align}
    H^\prime_N(\bm k)&= \beta \bm g_Z (\bm k) \cdot \bm \sigma 
    = \beta \sin k_x \sigma_z, \label{eq:ZeemanSOC}
\end{align}
as an example of the perturbation. This term is called Zeeman-type (or, Ising-type) SOC and is known to play an important role in two-dimensional transition metal dichalcogenides~\cite{Lu2015, Saito2016, Zhou2016,  Nakamura2017,Saito2016_review,He2018}. Although there is no evidence for this SOC in the $d$-wave superconductors to the best of our knowledge, it can be induced by using a substrate with lower symmetry. We will discuss this point later in Sec.~\ref{sec:discussion_exp}. 

Let us see the effect of the perturbation \eqref{eq:ZeemanSOC}. The quasiparticle spectra with this perturbation are shown in Figs.~\ref{fig:dwave}~(c), (d), and (f). Even with the Zeeman SOC, the gap nodes still exist without supercurrent [Fig.~\ref{fig:dwave}~(c)]. In contrast, with both finite Zeeman SOC and supercurrent, these nodes are gapped out [Figs.~\ref{fig:dwave}~(d) and (f)]. In this gapped phase, we calculate the energy spectrum with OBC and find the Majorana edge modes as shown in Fig.~\ref{fig:dwave}~(i). This suggests that a TSC is realized with supercurrent. As in the previous proposals of topological $d$-wave superconductors with breaking time-reversal symmetry~\cite{Daido2016, Takasan2017a}, the Chern number can be used for characterizing the TSC in our case. The Chen number is defined as
\begin{align}
    C \! = \frac{1}{2\pi i} \! \sum_{
    \substack{n=1,2,\\ i,j=x,y}} \int \! d\bm k(i \sigma_y)_{ij} \partial_{k_i} \!\! \bra{u_n(\bm k)}\partial_{k_j} \! \ket{u_n(\bm k)}. \label{eq:Chern}
\end{align}
The bands of $n=1,2$ are the states having the two lowest energy eigenstates for each momentum.
For numerical calculation, we use the method for the discretized Brillouin zone~\cite{Fukui2005}. The results of the Chern number are summarized in Fig.~\ref{fig:dwave}~(j). This figure shows that TSCs with $|C|=4$ are realized with infinitesimal supercurrent in the $x$-direction. This means that we can induce the TSC in $d$-wave superconductors with a certain perturbation by a very small supercurrent. This is one of the most important results of this study. We emphasize that the Zeeman-type SOC \eqref{eq:ZeemanSOC} is just an example of the perturbation breaking the symmetry protecting the nodes. It can be possible to find other perturbations.

\begin{figure}[t]
\includegraphics[width=8.5cm]{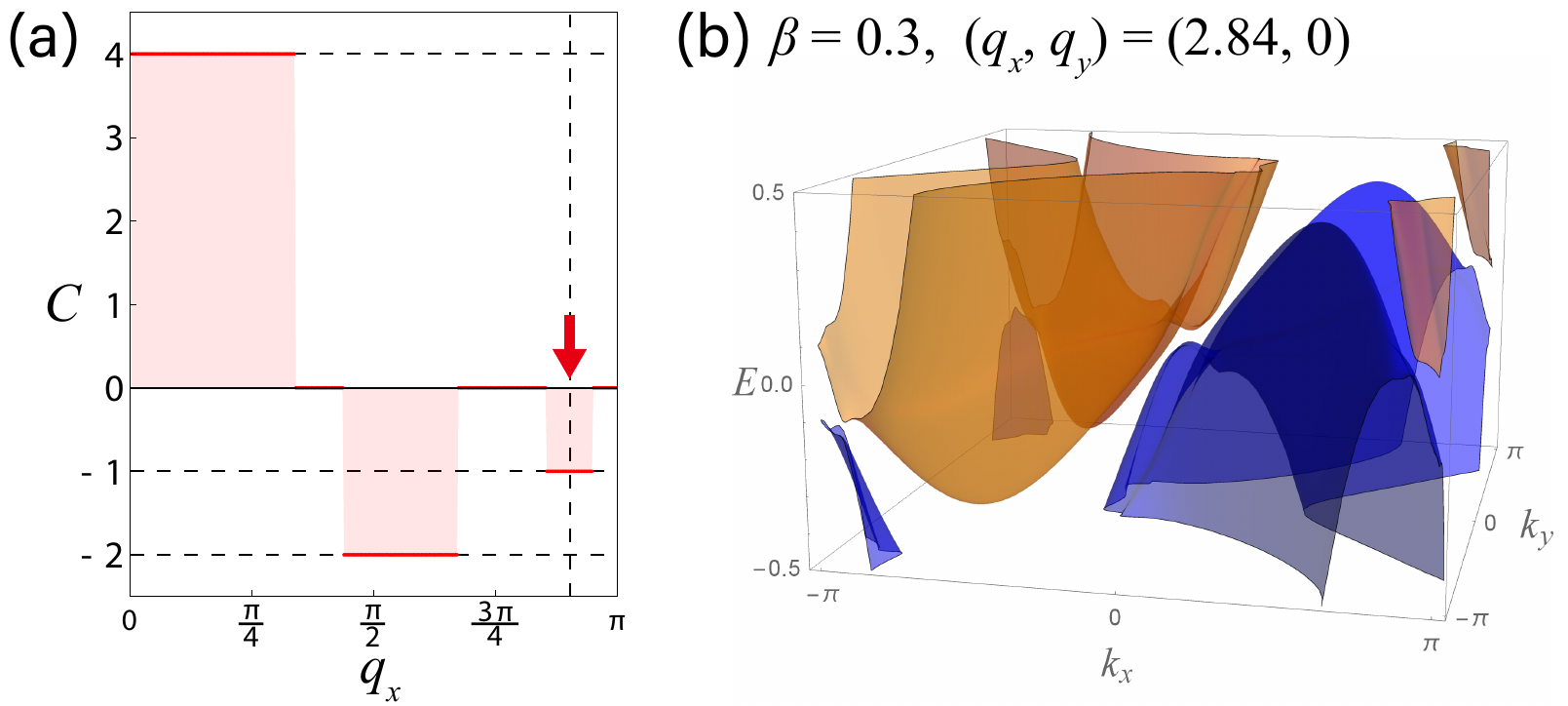}
\caption{Results of the $d$-wave superconductors [Eqs.~\eqref{eq:dwave_normal} and \eqref{eq:dwave_OP}] in the large-$\bm q$ regime. (a) Chern number for $0 < q_ x< \pi$ and $q_y=0$. (b) Quasiparticle spectrum for $\bm q$=(2.84, 0) denoted by the red arrow in the panel (a). 
The other parameters are the same as Fig.~\ref{fig:dwave}.
}
\label{fig:large_q}
\end{figure}

Before closing this section, we briefly mention the topological phases in the large-$\bm q$ regime, which is difficult to be achieved in current-driven superconductors but can be realized in intrinsically finite COM pairing states. With increasing the COM momentum, topological phase transitions occur again and several phases with the different Chern numbers appear. The change of the Chern number is summarized in Fig.~\ref{fig:large_q}~(a) and we can find a $C=-1$ phase for $2.69 \lesssim q_x \lesssim 2.98$ and $q_y=0$. This phase is interesting because, in the previous studies for TSCs in $d$-wave superconductors~\cite{Daido2016, Takasan2017a} and our Fig.~\ref{fig:dwave}~(j), there only appear even Chern numbers. The TSC with even Chern number can only host pairs of Majorana fermions, not a single one, which is useful for quantum computation~\cite{Alicea2012}. To realize a single Majorana fermion, we need a very large $q$, and thus the candidate may be the pair density wave state proposed for strongly correlated electron systems~\cite{Agterberg2020}. The relevance of our results in such finite COM pairing states will be discussed in Sec.~\ref{sec:discussion_finite_q}.

\section{Origin of the quantized Berry phase}
\label{sec:quantization}
In the results of $s$-wave and $d$-wave superconductors presented in Secs.~\ref{sec:s-wave} and \ref{sec:d-wave}, the point nodes appear in a wide range of the COM momentum $\bm q$ even under supercurrent. We have also shown that they are protected by the quantized Berry phase. 
Below, we explain that the protection can be understood from the classification of topological phases.

To understand the effect of supercurrent, we need to consider the spatial symmetry related to the current. For this purpose, we use the results of AZ+$\mathcal{I}$ classification, which gives the information of possible topological phases under the composite symmetry of Altland-Zirnbauer-type (onsite) symmetry and inversion symmetry~\cite{Bzdusek2017}.
Although the inversion symmetry is broken in our models of $s$-wave and $d$-wave superconductors, the twofold rotational symmetry $C_{2z}$ instead plays an important role since it transforms $\bm{k}$ to $-\bm{k}$ in two-dimensional systems, like the inversion symmetry. On any closed (one-dimensional) path in the Brillouin zone, the combined symmetry $TC_{2z}$ is preserved and squares to $+1$, even when supercurrent is applied. Based on Ref.~\cite{Bzdusek2017}, the symmetry configuration corresponds to the AI class with $Z_2$ topology, which is represented by the quantized Berry phase in Eq.~\eqref{eq:Berry}. That is why the point nodes can be characterized by the quantized Berry phase [Figs.~\ref{fig:swave}(h) and \ref{fig:dwave}(h)].
There are other types of phases in the periodic table, and it will be an interesting direction to study the other topological phase transitions and nontrivial nodal structures in current-driven superconductors based on the classification data.

\section{Discussion}
\label{sec:discussion}

\subsection{Experimental setup}
\label{sec:discussion_exp}
We discuss the experimental setup for the supercurrent-induced topological phase transitions. 
First, we consider the candidate materials. 
For our predictions in $s$-wave and $d$-wave superconductors, the Rashba SOC plays an essential role, and thus two-dimensional thin films of superconductors fabricated on substrates are good candidates~\cite{Saito2016_review}.
To realize fully-gapped TSCs with supercurrent in the $d$-wave case, we need to break the $T C_{2z}$ symmetry. In Sec.~\ref{sec:d-wave}, we have considered a perturbation \eqref{eq:ZeemanSOC}, which breaks the mirror symmetry in the $x$-direction. One possible way to break the mirror symmetry is to use twisted double-layered cuprate thin films~\cite{Can2021, Volkov2020}. Also, using a substrate lacking mirror symmetry might be another way. Since the symmetry is broken in materials with lower symmetry or with adding adequate perturbations, such as external fields, there are various possibilities. It is an important direction to search for good candidate materials and setups.

As for the application of supercurrent, we have to pay attention to the following two points. One is that supercurrent runs only near the surface of samples due to the Meissner effect~\cite{Tinkham_book}, though we assume in our theory that the finite COM momentum $\bm q$ arises uniformly in space. To realize the uniform current, we can use a narrow strip with a width of 0.1-1.0 $\mu$m order. For two-dimensional superconductors, the length scale of the current-carrying area is given by the Pearl length $\Lambda = 2 \lambda^2/d$ where $\lambda$ is the magnetic penetration depth and $d$ is the sample thickness~\cite{Pearl1964, Clem2011}. Using a sample narrower than the Pearl length $\Lambda$, the current distributes almost uniformly in the sample. Although the Pearl length is the order of 0.1-1~$\mu$m, it is possible to make a sample smaller than the length scale thanks to the development of the nanopatterning techniques for superconductors~\cite{Nawaz2013, Li2013, Sun2020}. Indeed, in experiments for cuprate superconductors~\cite{Nawaz2013} and iron-based superconductors~\cite{Li2013, Sun2020}, this type of sample is realized and the results support the spatially uniform current. 
Another way is to see the center region of samples, narrow but broader than the Pearl length~\cite{Nakamura2020}. Due to the incomplete Meissner effect, a small but finite current may almost uniformly flow in this region.

The other point is that the transition momentum $q_T$ must be smaller than the critical momentum $q_c$, which corresponds to the critical current $j_c$. To obtain the larger effect of supercurrent, systems with larger critical current are desirable. For this purpose, it is important to use thin, narrow, and clean samples in order to avoid the incursion of magnetic vortices. This is because the vortex flow induces a dissipative current, and $j_c$ typically becomes very small. Indeed, such samples avoiding vortices are used in experiments, and the large critical current, called depairing current~\cite{Tinkham_book}, is observed~\cite{Nawaz2013, Li2013, Sun2020}. However, even if we avoid the above difficulty, there are still problems. In particular, the depairing current for $s$-wave superconductors is determined by the Landau's criterion for superfluidity~\cite{Tinkham_book,Daido2021}, indicating that the gapless state is unstable. Thus, 
the $s$-wave case in Sec.~\ref{sec:s-wave} might be difficult to be realized, since the topological phase transition occurs at the gapless transitions, i.e., $q_T \simeq q_c$. Though, it is difficult to conclude it from the following reasons. (i) It is known that the depairing current can depend on the sample detail, such as the sample geometry~\cite{Clem2011, Hortensius2012, Nawaz2013}, and there might be a way to obtain a larger critical current. (ii) The $s$-wave pairing is the most widely realized in superconductors, and this point can be advantageous for searching the candidate materials. On the other hand, to avoid the problem of $q_T \simeq q_c$, the $d$-wave case is suitable. This is because the infinitesimal supercurrent realizes topological phase transition $q_T=0$ and thus $q_T < q_c$ is always satisfied under the supercurrent flow. For this case, we need to break the $TC_{2z}$ symmetry to realize the phase transition. 

We also discuss the experimental methods. The most direct evidence for the topological phase transition is the edge modes. To observe them, scanning tunnel microscope or angle-resolved photoemission spectroscopy work as useful methods. Also, indirect but strong evidence is to detect the change in the structure of the superconducting gap. 
To detect it, the transport or optical measurements, which can capture the information away from the Fermi level, are expected to be useful. This is because the change of the gap nodes does not occur at the Fermi energy due to the Doppler shift. In addition, the Bogoliubov Fermi surfaces (BFS) can appear under a supercurrent, and the thermodynamic properties are expected to be dominated by the BFS. Thus, we need to access the information of the excited state. 

\subsection{Relevance to finite COM pairing states}
\label{sec:discussion_finite_q}

We briefly comment on the finite COM pairing states. In this study, we consider the BdG Hamiltonian with a finite COM momentum $\bm q$ as a model for the current-driven superconductors and treat $\bm q$ as a controllable external field. The mean-field Hamiltonian is formally the same as the finite COM pairing states, spontaneously realized without using any external current source. 

Our mean-field Hamiltonian [Eq.~\eqref{eq:MFHam_1}] is the same as the FF superconductivity~\cite{Fulde1964}, where the Cooper pair has a single COM momentum and the inversion symmetry is spontaneously broken. Thus, when the FF superconductivity is realized and the COM momentum exceeds the transition momentum $q_T$, there occurs the topological phase transition. Since the COM momentum of the FF state is typically small in the order of inverse coherence length, this is a suitable platform for the $d$-wave superconductors, where $q_T$ is zero. For this direction, a good candidate can be CeCoIn$_5$, where the $d$-wave superconductivity is established~\cite{115_review} and the Fulde-Ferrel-Larkin-Ovchinnikov (FFLO) state is expected to be realized~\cite{Bianchi2003}.
Note that the single-$\bm{q}$ FF state is expected to be stabilized under the application of a small current to the double-$\bm{q}$ Larkin-Ovchinnikov state.
Another good candidate is odd-parity magnetic multipole systems.
Previous studies have shown that the FF state, rather than the LO state, coexisting with an odd-parity magnetic quadrupole order is stabilized without an external magnetic field in spin-orbit coupled superconductors~\cite{Sumita2016, Sumita2017}.
Since the COM momentum can be tuned by the magnitude of the magnetic moments, the odd-parity magnetic multipole systems could also be a good platform to realize the supercurrent-induced topological phase transition.

To realize a large COM momentum beyond the Laudau's criterion, the pair-density wave (PDW) superconductors~\cite{Agterberg2020} are promising. The PDW state is essentially the same as the FFLO state except for one difference that the pairing amplitude modulates in the scale of the lattice constant, while the length scale of the modulation in the FFLO state is the order of the coherence length. The short length scale means the large COM momentum. The appearance of the PDW states has been discussed in strongly correlated materials including cuprates and CeCoIn$_5$~\cite{Agterberg2020}. Investigating the topologically non-trivial states predicted in our study, such as the $|C|=1$ state in the large-$\bm q$ regime, will be interesting.

\section{Conclusion and Outlook}
\label{sec:conclusion}

In this paper, we have studied the topological phase transitions in the quasiparticle spectrum of the current-driven superconductors. Taking the effect of supercurrent into account as the COM momentum $\bm q$, we have studied three prototypical examples, the Kitaev chain, $s$-wave superconductors, and $d$-wave superconductors. The results show that all the models exhibit topological phase transitions induced by supercurrent. This suggests that the mechanism of the supercurrent-induced topological phase transition is generic and widely applicable to various materials. We have also found that the protection of the gap nodes under a finite current is related to the composite symmetry $T C_{2z}$, and this tells us what kind of perturbation is useful for controlling the structure of the nodes with supercurrent. In addition, we have addressed the experimental setup to verify our prediction and the relevance to the finite COM pairing states.

In addition to the future issues we already mentioned in Sec~\ref{sec:discussion}, we have two important future directions. One is to extend our analysis to a wide range of materials. While we have studied one- and two-dimensional systems, it is important to study other examples, such as multilayered systems or three-dimensional materials, since they are also important platforms to study novel superconductivity. The other is to propose a setup more feasible in order to realize in experiments. As we discussed in Sec.~\ref{sec:discussion_exp}, there still exist several difficulties, although we can consider several possible setup for the  experimental observation of the current-induced topological phase transitions. It is important to search for a new setup for overcoming the difficulties. We hope that our work opens new research directions to control the topological phases of matter and to realize a novel TSC.

\begin{acknowledgments}
We thank Akito Daido, Joel E. Moore, and Micha{\l} Papaj for valuable discussions. K.T. is supported by the U.S. Department of Energy (DOE), Office of Science, Basic Energy Sciences (BES), under Contract No. AC02-05CH11231 within the Ultrafast Materials Science Program (KC2203). S.S. is supported by JST CREST Grant No. JPMJCR19T2. Y.Y. is supported by JSPS KAKENHI (Grants No. JP18H05227, No. JP18H01178, No. 20H05159) and SPIRITS 2020 of Kyoto University.
\end{acknowledgments}

\appendix

\section{Relation between the winding number and the Berry phase}
\label{sec:wind_Berry}

In this Appendix, we show that there exists a relation between the winding number [Eq.~\eqref{eq:wind_num}] and the Berry phase [Eq.~\eqref{eq:Berry}], when both indices are well-defined (i.e., the COM momentum $\bm{q}$ is parallel to the $x$- or $y$-direction).
The relation is given by
\begin{equation}
 \frac{\gamma}{\pi} = w(k_x) \pmod{2\pi},
 \label{eq:wind_Berry}
\end{equation}
where the Berry phase in the LHS and the winding number in the RHS are defined on the same closed path $C$, namely, $C = \{(0, k_y) | -\pi \leq k_y < \pi\}$ or $\{(\pi, k_y) | -\pi \leq k_y < \pi\}$.
In the following, we prove Eq.~\eqref{eq:wind_Berry} based on discussions in Ref.~\cite{Ryu2010}.

As mentioned in Sec.~\ref{sec:s-wave}, the chiral symmetry $\Gamma$ ($\Gamma^2 = +1$) is preserved on the path $C$:
\begin{equation}
 \{H_{\text{BdG}}(\bm{k}), \Gamma\} = 0, \quad \bm{k} \in C.
\end{equation}
Therefore, the spectrum of the BdG Hamiltonian for $\bm{k} \in C$ is symmetrical about the zero energy, which indicates the number of the positive eigenvalues is the same as that of the negative ones: $N_+ = N_- =: N$.
In the model of $s$-wave superconductors, $N$ is equal to $2$.
Let $\{\ket{u^\pm_{\hat{n}}(\bm{k})}\}_{\hat{n}=1}^{N}$ be a set of the positive (negative) eigenstates of the BdG Hamiltonian.
By using this, we define a $Q$-matrix as
\begin{equation}
 Q(\bm{k}) := \sum_{\hat{n}=1}^{N} \left[ \ket{u^+_{\hat{n}}(\bm{k})}\!\bra{u^+_{\hat{n}}(\bm{k})} - \ket{u^-_{\hat{n}}(\bm{k})}\!\bra{u^-_{\hat{n}}(\bm{k})} \right],
\end{equation}
which means spectral flattening of the Hamiltonian.
Let us consider a basis set diagonalizing the chiral symmetry,
\begin{equation}
 \Gamma = \text{diag}(\overbrace{+1, \dots, +1}^{N}, \, \overbrace{-1, \dots, -1}^{N}).
\end{equation}
Then the $Q$-matrix becomes off-diagonal,
\begin{equation}
 Q(\bm{k}) =
 \begin{pmatrix}
  0 & q(\bm{k}) \\
  q^\dagger(\bm{k}) & 0
 \end{pmatrix}, \quad
 q(\bm{k}) \in \text{U}(N),
\end{equation}
which corresponds to Eq.~\eqref{eq:off-diagonal_BdG}.

Now we consider the Berry connection of the chiral-symmetric Hamiltonian.
Since the equation
\begin{equation}
 Q(\bm{k}) \ket{u^\pm_{\hat{n}}(\bm{k})} = \pm \ket{u^\pm_{\hat{n}}(\bm{k})},
\end{equation}
is satisfied, the Bloch state is represented by, after some algebra,
\begin{equation}
 \ket{u^\pm_{\hat{n}}(\bm{k})} = \frac{1}{\sqrt{2}}
 \begin{pmatrix}
  \pm q(\bm{k}) \eta^\pm_{\hat{n}} \\
  \eta^\pm_{\hat{n}}
 \end{pmatrix}.
\end{equation}
Therefore, the $N$-dimensional space spanned by the occupied states $\{\ket{u^-_{\hat{n}}(\bm{k})}\}_{\hat{n}=1}^{N}$ can be obtained by $N$ orthonormal vectors $\eta^\pm_{\hat{n}}$.
Although $\eta^\pm_{\hat{n}}$ in general depends on $\bm{k}$, we can always choose $\bm{k}$-\textit{independent} orthonormal vectors; for convenience, we choose $(\eta^\pm_{\hat{n}})_{\hat{m}} = \delta_{\hat{n}\hat{m}}$ in the following discussions.
In this gauge, the Berry connection is calculated as
\begin{align}
 \bm{A}^\pm_{\hat{n}}(\bm{k}) &:= \braket{u^\pm_{\hat{n}}(\bm{k}) | \nabla_{\bm{k}} u^\pm_{\hat{n}}(\bm{k})} \notag \\
 &= \frac{1}{2} \left[ \braket{\eta^\pm_{\hat{n}} q^\dagger(\bm{k}) | \nabla_{\bm{k}} | q(\bm{k}) \eta^\pm_{\hat{n}}} + \braket{\eta^\pm_{\hat{n}} | \nabla_{\bm{k}} | \eta^\pm_{\hat{n}}} \right] \notag \\
 &= \frac{1}{2} [q^\dagger(\bm{k}) \nabla_{\bm{k}} q(\bm{k})]_{\hat{n}\hat{n}}.
\end{align}
The Berry phase is finally calculated by
\begin{align}
 \gamma &= \frac{1}{i} \sum_{\hat{n}} \oint_{C} d\bm{k} \cdot \bm{A}^-_{\hat{n}}(\bm{k}) \notag \\
 &= \frac{1}{2i} \int_{-\pi}^{\pi} \!\! dk_y \text{tr}[q^\dagger(\bm{k}) \partial_{k_y} q(\bm{k})] \notag \\
 &= \frac{1}{2} \im\left[ \int_{-\pi}^{\pi}\!\!dk_y \partial_{k_y} \! \ln \det \hat{q}(\bm k) \right] \notag \\
 &= \pi w(k_x) \pmod{2\pi}.
\end{align}
Although we have made the specific choice of the gauge in the above calculations, the Berry phase modulo $2\pi$ is invariant under gauge transformation.
Therefore, we have proved that the relation~\eqref{eq:wind_Berry} holds for any gauge choice.

\section{Perturbation theory around the nodes}
\label{sec:perterbation}

We evaluate the current-induced gap around the nodes in $d$-wave superconductors in a perturbative way, using the result obtained in Ref.~\cite{Daido2016}. First, we use the following Nambu basis,
\begin{align}
    \bm \tilde{\Psi}_{\bm k;\bm q}=(c_{\bm k+\bm q, \uparrow}, c_{\bm k+\bm q, \downarrow},c^\dagger_{-\bm k+\bm q, \downarrow}, -c^\dagger_{-\bm k+\bm q, \uparrow})^T. \label{eq:different_basis}
\end{align}
Using this basis, the mean-field Hamiltonian Eq.~\eqref{eq:MFHam_2} is written as $H=(1/2)\sum_{\bm k} \bm \tilde{\Psi}^\dagger_{\bm k; \bm q} \tilde{H}_\mathrm{BdG}(\bm k; \bm q) \bm \tilde{\Psi}_{\bm k; \bm q}$ with
\begin{align}
    \tilde{H}_\mathrm{BdG}(\bm k; \bm q) \! = \!
    \begin{pmatrix}
    H_N(\bm k+\bm q) & \tilde{\Delta}(\bm k) \\
    \tilde{\Delta}^\dagger(\bm k) & -T H_N(\bm k+\bm q) T^{-1}
    \end{pmatrix},
\end{align}
where $\tilde{\Delta}(\bm k)= - \Delta(\bm k) (i\sigma_y)$ and $T$ is the time-reversal operator acting as $T H_N(\bm k+\bm q) T^{-1}= i\sigma_y H^*_N(-\bm k+\bm q) (-i\sigma_y)$. $\bm \sigma=(\sigma_x, \sigma_y, \sigma_z)$ are the Pauli matrices in the spin space.

Next, we expand $H_N(\bm k)$ and $\tilde{\Delta}(\bm k)$ as $H_N(\bm k)=\xi(\bm k)\sigma_0 + \bm g(\bm k) \cdot \bm \sigma$ and $\tilde{\Delta}(\bm k)=\psi(\bm k)\sigma_0 + \bm d(\bm k)\cdot \bm \sigma$, with the identity matrix in the spin space $\sigma_0$.  We assume $\xi_{\bm q}(\bm k)$ and $\bm g_{\bm q}(\bm k)$ are parity-even and parity-odd, respectively. This assumption is valid when the time-reversal symmetry exists before applying supercurrent, which is the case we consider in this study. Then, we obtain the BdG Hamiltonian up to the first order of $q=|\bm q|$ as
\begin{align}
\tilde{H}_\mathrm{BdG}(\bm k;\bm q)&= H_N(\bm k) \otimes \tau_z  + \tilde{\Delta}(\bm k) \otimes \tau_x \nonumber + \xi_{\bm q}^\prime(\bm k) \sigma_0 \otimes \tau_0  \\
&\quad + \bm g_{\bm q}^\prime(\bm k) \cdot \bm \sigma \otimes \tau_0 + O(q^2), \label{eq:tildeBdG_with_q}
\end{align} 
where
\begin{align}
    \xi_{\bm q}^\prime(\bm k)  = \! \! \sum_{i=x,y} \! \! q_i \frac{\partial \xi(\bm k)}{\partial k_i},\quad
    \bm g_{\bm q}^\prime(\bm k)= \! \! \sum_{i=x,y} \! \! q_i \frac{\partial \bm g(\bm k)}{\partial k_i}.
\end{align}
Here, $\tau_0$ and $\bm \tau=(\tau_x, \tau_y, \tau_z)$ are the identity matrix and the Pauli matrices in the Nambu space.
Note that $\xi_{\bm q}^\prime(\bm k)$ and $\bm g_{\bm q}^\prime(\bm k)$ are parity-odd and parity-even, respectively. In Eq.~\eqref{eq:tildeBdG_with_q}, the third and fourth terms represent the leading effect of supercurrent. The third term is proportional to the $4 \times 4$ identity matrix and gives the momentum-dependent energy shift. This term deforms the energy spectrum but does not affect the structure of nodes. In contrast, the fourth term is spin-dependent and can give a non-trivial effect on the gap nodes. 

For understanding the effect of supercurrent, it is useful to compare it with the effect of the Zeeman magnetic field. Using the Nambu basis \eqref{eq:different_basis}, the BdG Hamiltonian under the Zeeman field $\bm h$ is given by 
\begin{align}
\tilde{H}_\mathrm{BdG}(\bm k)& \! = \! H_N(\bm k) \otimes \tau_z  + \tilde{\Delta}(\bm k) \otimes \tau_x + \bm h \cdot \bm \sigma \otimes \tau_0. \label{eq:tildeBdG_with_h}
\end{align}
We can see that the third term in Eq.~\eqref{eq:tildeBdG_with_h} with replacement $\bm h \to \bm g_{\bm q}^\prime(\bm k)$ is the same as the fourth term in Eq.~\eqref{eq:tildeBdG_with_q}. Using this correspondence, we apply the results for the Zeeman magnetic field obtained in the previous works. In Ref.~\cite{Daido2016}, the quasiparticle spectrum of the BdG Hamiltonian~\eqref{eq:tildeBdG_with_h} is obtained based on the perturbation theory, and we use it to evaluate the current-induced gap. 
As a result with considering the energy shift by $\xi_{\bm q}^\prime(\bm k)$ in Eq.~\eqref{eq:tildeBdG_with_q}, the quasiparticle spectrum under supercurrent is given as
\begin{align}
    \calE^{(+)}_{\bm q, \pm} (\bm k)\!  &= \! \xi_{\bm q}^{\prime}(\bm k) \! - \! \bm g_{\bm q}^{\prime}(\bm k)\! \cdot \! \hat{\bm g}(\bm k) \! \pm \! \sqrt{E^{(+)}\!(\bm k)^2+|\Delta^{(+)}_{\bm q}\!(\bm k)|^2}, \label{eq:pert_quasiparticle_spectrum_+} \\
    \calE^{(-)}_{\bm q, \pm} (\bm k)\!  &= \! \xi_{\bm q}^{\prime}(\bm k) \! + \! \bm g_{\bm q}^{\prime}(\bm k)\! \cdot \! \hat{\bm g}(\bm k) \! \pm \! \sqrt{E^{(-)}\!(\bm k)^2+|\Delta^{(-)}_{\bm q}\!(\bm k)|^2}, \label{eq:pert_quasiparticle_spectrum_-}
\end{align}
where
\begin{align}
    E^{(\pm)}(\bm k)&=\xi(\bm k)\pm |\bm g(\bm k)|, \quad \hat{\bm g}(\bm k)=\bm g(\bm k)/|\bm g(\bm k)|,  \\
    i\Delta^{(\pm)}_{\bm q}(\bm k)&= \psi(\bm k)\pm \bm d(\bm k)\cdot\hat{\bm g}(\bm k) + i\frac{\bm g^\prime_{\bm q}(\bm k) \cdot \hat{\bm g}(\bm k)\times \bm d(\bm k)}
    {|\bm g(\bm k)|}. \label{eq:delta_pm}
\end{align}
Note that there are four bands $ \calE^{(+)}_{\bm q, +}(\bm k)$, $\calE^{(+)}_{\bm q, -}(\bm k)$, $\calE^{(-)}_{\bm q, +}(\bm k)$, and $\calE^{(-)}_{\bm q, -}(\bm k)$, since our BdG Hamiltonian is a $4 \times 4$ matrix. These formulae are derived based on these assumptions,
\begin{align}
    &|\psi(\bm k)| \ll |\bm g (\bm k)|, \quad
    |\bm d(\bm k)| \ll |\bm g (\bm k)|, \label{eq:pert_condition_1} \\
    &\qquad \qquad |\bm g^\prime_{\bm q}(\bm k)| \ll |\bm g (\bm k)|. \label{eq:pert_condition_2}
\end{align}
When the SOC gives an essential effect on the superconducting gap, the two assumptions \eqref{eq:pert_condition_1} are valid except for the zeros of $\bm g(\bm k)$. Thus, it is expected to be satisfied near the gap nodes in general. The assumption \eqref{eq:pert_condition_2} is valid with a sufficiently small supercurrent, which is nothing but the case we are considering here.

We want to evaluate the current-induced gap around the nodes. These nodes are characterized by the momenta satisfying $\calE^{(\pm)}_{\bm q= \bm 0, +}(\bm k_0)-\calE^{(\pm)}_{\bm q= \bm 0, -}(\bm k_0)=0$. With Eqs.~\eqref{eq:pert_quasiparticle_spectrum_+} and \eqref{eq:pert_quasiparticle_spectrum_-}, we can show
\begin{align}
    E^{(\pm)}(\bm k_0)=0, \quad
    \psi(\bm k_0)\pm \bm d(\bm k_0)\cdot\hat{\bm g}(\bm k_0)=0. \label{eq:node_condition}
\end{align}
Using the momenta of the nodes $\bm k_0$, the current-induced gap is defined as $\Delta^{(\pm)}_\mathrm{node}(\bm q) = |\calE^{(\pm)}_{\bm q, +}(\bm k_0)-\calE^{(\pm)}_{\bm q, -}(\bm k_0)|$. Using the relations in Eq.~\eqref{eq:node_condition}, we can evaluate $\Delta_\mathrm{node}(\bm q)=\Delta^{(+)}_\mathrm{node}(\bm q)=\Delta^{(-)}_\mathrm{node}(\bm q)$ as
\begin{align}
    \Delta_\mathrm{node}(\bm q) = 2 \left| \frac{\bm g^\prime_{\bm q}(\bm k_0) \cdot \hat{\bm g}(\bm k_0)\times \bm d(\bm k_0)}
    {\bm g(\bm k_0)} \right| + O(q^2).
\end{align}
This is the formula \eqref{eq:current_induced_gap} in the main text.

\bibliographystyle{apsrev4-2}
\bibliography{ref.bib}

\begin{thebibliography}{87}
\bibitem [{foo()}]{footnote1}%
  \BibitemOpen
  \href@noop {} {}\bibinfo {note} {This effect is not considered in the
  $s$-wave case in Sec.~\ref{sec:s-wave} because it gives no significant
  effects on the qualitative results.}\BibitemShut {Stop}%
\end{thebibliography}

\end{document}